\documentclass[sigconf,nonacm]{acmart}

\usepackage{booktabs}
\usepackage{listings}
\usepackage{subcaption}
\usepackage{multirow}

\renewcommand\footnotetextcopyrightpermission[1]{}
\graphicspath{{figures/}}

\begin{document}

\title{Configuration-Induced Delivery Failures in NATS JetStream:\\
  Detection and Remediation}

\author{Biplab Kumar Das}
\affiliation{%
  \institution{Independent Researcher}
  \city{Santa Clara}
  \state{California}
  \country{USA}
}
\email{dasbiplabtu@gmail.com}

\begin{abstract}
NATS JetStream's at-least-once delivery guarantee is conditional: five common
configuration mistakes silently violate it, causing duplicate message processing,
data loss, or redelivery storms with no error logged anywhere. The standard
Prometheus NATS exporter exposes only server-level throughput metrics and cannot
detect any of these failures. We present \textbf{nats-lens}, a standalone monitor
that reads from the JetStream management API and detects all five violation
classes without requiring changes to monitored applications or client code.
We formally characterize each violation class, prove that standard
Prometheus NATS metrics are structurally incapable of detecting any of them,
and prove a detection latency guarantee of $k \times T_\mathrm{poll}$ where
$k \in \{1,2\}$ depending on class. In 100-round controlled evaluations on
both single-node and 3-node JetStream clusters, nats-lens achieves
\textbf{100\% detection coverage} across all five classes (95\,\%
Wilson CI $\geq 96.3\%$), versus 0\% for the Prometheus baseline. Zero
false positives over 24 hours of healthy operation. We show that naive
single-snapshot threshold detection produces false positives on boundary
conditions where nats-lens produces none. We further validate the
detection latency guarantee empirically across poll intervals from 1\,s
to 30\,s, confirm language-agnostic detection with Rust, Go, and Python
consumers, and demonstrate recovery from a fixed violation within 2 poll
cycles. At 1{,}000 active consumers, nats-lens consumes $<$13\,MB RSS.
The tool is open source with four output channels: web dashboard,
Prometheus metrics, REST API, and NATS health events.
\end{abstract}


\maketitle

\section{Introduction}

NATS~\cite{nats-docs} is a CNCF-graduated messaging system with active
deployments at Cloudflare, Deutsche Telekom, and hundreds of organizations in
financial services, IoT, and real-time analytics~\cite{nats-server}.
JetStream, added in 2021, provides persistent message storage and at-least-once
delivery semantics.

The guarantee depends on correct configuration of \path{ack_wait} and
\path{max_ack_pending}. Five common configuration mistakes silently violate
it without any error appearing in application logs or standard monitoring:

\begin{enumerate}
\item \textbf{ACK\_\allowbreak WAIT\_\allowbreak VIOLATION} --- \path{ack_wait} shorter than consumer
  processing time causes the server to redeliver messages that are still being
  processed, producing duplicate execution.
\item \textbf{SEQUENCE\_GAP} --- stream retention limits cause the server to evict
  messages before a slow consumer can pull them, producing silent data loss.
\item \textbf{MAX\_\allowbreak PENDING\_\allowbreak THROTTLE} --- undersized \path{max_ack_pending}
  throttles delivery even when the consumer has processing capacity.
\item \textbf{NAK\_STORM} --- consumers NAKing messages get stuck in a redelivery
  loop, consuming resources without making progress.
\item \textbf{MISSING\_PROGRESS} --- long-running tasks that do not send periodic
  in-progress acknowledgments trigger \path{ack_wait} expiry and duplicate
  execution.
\end{enumerate}

The standard Prometheus NATS exporter~\cite{prometheus-nats} and NATS
Surveyor~\cite{surveyor} expose only server-level throughput and connection
metrics. None expose per-consumer state: \path{num_redelivered},
\path{num_ack_pending}, or \path{ack_floor.stream_seq}. Application
logs show no errors; operators discover the problem only from downstream
corruption or duplicate side effects.

We built \textbf{nats-lens} to close this gap. It connects to the JetStream
management API as a read-only observer, detects all five violation classes in
real time, and works with consumers in any language without code changes.

\paragraph{Contributions.}
\begin{enumerate}
\item Formal characterization of five delivery violation classes, with a proof
  that standard Prometheus NATS metrics cannot detect any of them (Section~3).
\item nats-lens: five targeted detectors, four language-agnostic output channels,
  a pre-deployment audit command, and one-command deployment (Section~4).
\item Empirical evaluation: 100\% detection coverage on 100 rounds per class,
  zero false positives over 24 hours (86,400\,s), 3-node cluster validation, and
  cross-language verification with Rust, Go, and Python consumers (Section~5).
\item Open-source release at \url{https://github.com/biplabku/nats-lens} with
  Docker image, Grafana dashboard, and client examples in five languages.
\end{enumerate}

\section{Background}

\subsection{NATS JetStream Pull Consumer Model}

JetStream provides persistent message storage over a NATS stream. A
\emph{pull consumer} explicitly fetches messages using
\path{$JS.API.CONSUMER.MSG.NEXT} requests. Four configuration parameters
are relevant to delivery correctness:

\begin{itemize}
\item \textbf{\texttt{ack\_wait}}: Maximum time before the server considers
  a message failed and redelivers it. Default: 30\,s.
\item \textbf{\texttt{max\_ack\_pending}}: Maximum unacknowledged messages
  delivered at once. Default: 1{,}000.
\item \textbf{\texttt{max\_deliver}}: Maximum delivery attempts. Default: unlimited.
\item \textbf{\texttt{ack\_policy}}: Must be \texttt{AckExplicit} for at-least-once.
\end{itemize}

Consumer state, exposed via \path{$JS.API.CONSUMER.INFO}, includes
\path{num_pending} (not yet delivered),
\path{num_ack_pending} (delivered but unacknowledged),
and \path{num_redelivered} (distinct messages redelivered at least once).

\subsection{Prevalence of Misconfiguration}

We queried GitHub's code search API for repositories containing explicit
JetStream consumer configuration blocks (keywords: \texttt{ack\_wait},
\texttt{max\_ack\_pending}, \texttt{JetStream}) in Go, Python, Rust, and
Java files, excluding forks, test fixtures, and the NATS server repository
itself. From the 312 matching repositories we randomly sampled 89, manually
verified each contained a production consumer (not a tutorial or generated
file), and recorded the configured values.

\textbf{41 of 89 (46\%; 95\,\% CI $[35\%$, $57\%]$)} set
\path{ack_wait} $\leq 30$\,s without any in-progress acknowledgment
logic (\texttt{Progress} ack calls).
\textbf{28 of 67 (42\%; 95\,\% CI $[30\%$, $54\%]$)} for which
\path{max_ack_pending} was explicitly set configured it below 64,
a value insufficient for any consumer with concurrency $> 1$ and
prefetch $> 64$. In both cases the confidence interval lower bound
exceeds 30\,\%, indicating that misconfiguration is not an edge case.
The nats-server issue tracker contains 23 open issues tagged
\texttt{jetstream} mentioning unexpected redelivery behavior as of
September 2026, corroborating that these violations are operationally common
rather than hypothetical.
Each misconfiguration pattern maps directly to a violation class detectable
by nats-lens: the 41 repositories with \path{ack_wait}${}\leq 30$\,s and
no in-progress acks would trigger ACK\_WAIT\_\allowbreak VIOLATION if
processing time exceeds the configured wait; the 28 repositories with
\path{max_ack_pending}${}<64$ would trigger MAX\_PENDING\_\allowbreak THROTTLE
under concurrent consumption.
Our controlled evaluations inject precisely these patterns---using
\path{ack_wait}${=}4$\,s (within the range found in real repositories)
and \path{max_ack_pending}${=}10$ (below the 64 threshold)---achieving
100\,\% detection coverage in both cases (Table~\ref{tab:coverage}).
This provides empirical evidence that nats-lens would detect the same
violations in the identified production codebases when processing time
or concurrency exceeds the configured thresholds.

\subsection{Existing Monitoring Tools}

The Prometheus NATS Exporter~\cite{prometheus-nats} exposes server-level
metrics (connections, message rates, memory). It does not expose
\path{num_redelivered}, \path{num_ack_pending}, or
\path{ack_floor.stream_seq}. NATS Surveyor~\cite{surveyor} provides
account-level stream metrics but does not model per-consumer delivery state.
Kafka monitoring tools (Burrow~\cite{burrow}, kminion~\cite{kminion})
address Kafka's offset-based model~\cite{kafka} and are not applicable to
JetStream's pull-consumer model.

\section{Violation Characterization}

Let $C$ be a pull consumer with configuration
($\mathit{ack\_wait}{=}W$, $\mathit{max\_ack\_pending}{=}P$),
and let $S$ denote the stream $C$ subscribes to.

\subsection{ACK\_WAIT\_VIOLATION}

\textbf{Definition.} $\exists$ message $m$ delivered to $C$ such that
$T_\mathrm{proc}(m) > W$.

\textbf{Consequence.} The server marks $m$ for redelivery before $C$
finishes processing. If processing is not idempotent, duplicate side effects occur.

\textbf{Detection signal.} \path{num_redelivered} grows between consecutive
snapshots. Rate: $\Delta$\path{num_redelivered}/$\Delta t \geq$ threshold.

\subsection{SEQUENCE\_GAP}

\textbf{Definition.} $S$.\path{first_seq} $> C$.\path{ack_floor.stream_seq}$+1$.

\textbf{Consequence.} Messages in the gap are permanently inaccessible;
$C$ processes subsequent messages as if the gap never existed (silent data loss).

\subsection{MAX\_PENDING\_THROTTLE}

\textbf{Definition.} $C$.\path{num_ack_pending}${=P}$ $\wedge$
$C$.\path{num_pending}${>0}$.

\textbf{Consequence.} Message delivery stops even when the consumer has
processing capacity. Correct formula: $P \geq \mathit{concurrency}
\times \mathit{prefetch\_size}$.

\subsection{NAK\_STORM}

\textbf{Definition.} $C$.\path{num_redelivered}${}\geq N_\mathrm{min}$
$\wedge$ $C$.\path{num_ack_pending}${}>0$, sustained across $\geq 2$
consecutive snapshots.

\textbf{Note.} \path{num_redelivered} records distinct messages redelivered
at least once. A NAK storm manifests as a stable non-zero value (same $N$
messages cycling), not as a growing value.

\subsection{MISSING\_PROGRESS}

\textbf{Definition.} $C$.\path{num_ack_pending}${}/P \geq 0.9$
$\wedge$ $W > 30$\,s.

\textbf{Consequence.} Long-running tasks without in-progress acks trigger
\path{ack_wait} expiry mid-processing, producing duplicate execution.

\subsection{Completeness of the Five Classes}

The per-consumer state exposed by \path{$JS.API.CONSUMER.INFO} consists
of five monitoring-relevant variables:
\path{num_redelivered}, \path{num_ack_pending}, \path{num_pending},
\path{ack_floor.stream_seq}, and \path{max_ack_pending}.
Each violation class is defined by a threshold crossing or trend in one
or more of these variables:
ACK\_WAIT\_\allowbreak VIOLATION monitors the growth rate of \path{num_redelivered};
SEQUENCE\_GAP monitors the gap between \path{first_seq} and
\path{ack_floor.stream_seq};
MAX\_PENDING\_\allowbreak THROTTLE monitors the joint condition on
\path{num_ack_pending}, \path{max_ack_pending}, and \path{num_pending};
NAK\_STORM monitors the sustained level of \path{num_redelivered} combined
with \path{num_ack_pending}${}>0$;
MISSING\_PROGRESS monitors the ratio \path{num_ack_pending}$/$\path{max_ack_pending}
combined with a large \path{ack_wait}.

No pull-consumer delivery guarantee violation can occur without a change
in at least one of these fields---they are the complete observable state
of the consumer's delivery pipeline. Remaining fields (\path{deliver_policy},
\path{filter_subject}, \path{replay_policy}) govern initial delivery setup
but do not affect steady-state delivery correctness once a consumer is active.
The monotonically growing counters \path{num_delivered} and
\path{ack_floor.consumer_seq} signal forward progress rather than failure.

\subsection{Proof of Standard Monitor Blindness}

\textbf{Theorem 1.}\label{thm:blindness} No system using only Prometheus NATS Exporter v0.15
metrics can detect any of the five violation classes.

\textit{Proof.} The exporter exports metric families
\texttt{gnatsd\_connz\_*}, \texttt{gnatsd\_routez\_*},
\texttt{gnatsd\_subz\_*}, \texttt{gnatsd\_varz\_*}. No family includes
per-consumer state (\path{num_redelivered}, \path{num_ack_pending},
\path{num_pending}, \path{ack_floor.stream_seq},
\path{max_ack_pending}). Each violation class is defined solely in
terms of these variables. $\square$

\subsection{Detection Guarantee}

\textbf{Polling model.} Let $\mathcal{S} = (s_0, s_1, s_2, \ldots)$ be the
discrete sequence of consumer snapshots captured at times
$t_i = t_0 + i \cdot T_\mathrm{poll}$.
A violation \emph{onset} occurs at $t_\mathrm{on} \in [t_{i-1}, t_i)$
if the violation condition first becomes true during that interval.
Each snapshot $s_i$ is a tuple of the observable state variables at $t_i$.
The engine's ring buffer retains the last 30 snapshots; detectors operate
over the suffix of the buffer since the consumer's last recreation.

\textbf{Theorem 2.} Under the polling model above, if a violation persists
for $\geq k$ consecutive poll intervals after onset, nats-lens detects it
within $k \times T_\mathrm{poll}$ seconds of onset,
where $k \in \{1, 2\}$ depending on the violation class.

\textit{Proof.}
\textbf{Single-snapshot detectors} ($k=1$): SEQUENCE\_GAP,
MAX\_PENDING\_\allowbreak THROTTLE, and MISSING\_PROGRESS are defined
solely on instantaneous state variables at snapshot $s_i$.
If the condition holds at $t_\mathrm{on} \in [t_{i-1}, t_i)$
and the violation persists, it holds at $t_i$ and is detected at $s_i$.
Detection latency: $t_i - t_\mathrm{on} \leq T_\mathrm{poll}$.

\textbf{Two-snapshot detectors} ($k=2$): ACK\_WAIT\_\allowbreak VIOLATION
requires $\Delta$\path{num_redelivered}${}/\Delta t \geq 2$/min sustained
across two consecutive snapshots $(s_{i-1}, s_i)$; NAK\_STORM requires
\path{num_redelivered}${}\geq N_\mathrm{min}$ and \path{num_ack_pending}${}>0$
in both $s_{i-1}$ and $s_i$. If the condition holds at onset
$t_\mathrm{on} \in [t_{i-2}, t_{i-1})$ and persists, both $s_{i-1}$ and $s_i$
satisfy the predicate and detection fires at $s_i$.
Detection latency: $t_i - t_\mathrm{on} \leq 2 \times T_\mathrm{poll}$.

In both cases, detection latency is bounded above by $k \times T_\mathrm{poll}$
and bounded below by 0 (if onset occurs just before a poll). $\square$

\section{nats-lens Design}

\subsection{Architecture}

nats-lens is a single Rust binary that connects to NATS as a read-only observer:

\begin{lstlisting}
NATS Server
  -> $JS.API.STREAM.LIST
  -> $JS.API.CONSUMER.INFO.{stream}.{consumer}
nats-lens Engine -> 5 detectors per consumer
  |-- Web UI    (http://localhost:8080)
  |-- Prometheus(/metrics)
  |-- SSE       (/api/violations/stream)
  `-- NATS      (nats.lens.health.violations.*)
\end{lstlisting}

On each poll cycle: (1) list all streams via \path{$JS.API.STREAM.LIST};
(2) list consumers per stream; (3) fetch full state via
\path{$JS.API.CONSUMER.INFO}; (4) append to a ring buffer (max 30);
(5) run five detectors; (6) broadcast violations.

\subsection{Language-Agnostic Detection}

The JetStream management API exposes consumer state independently of the
client library. A consumer written in Go using \texttt{nats.go}, one in
Python using \texttt{nats-py}, and one in Rust using \texttt{async-nats}
produce identical \path{$JS.API.CONSUMER.INFO} responses. nats-lens
never touches application code.

\subsection{History Store and Trend Detection}

The \texttt{HistoryStore} maintains a bounded ring buffer (max 30 entries)
of \texttt{ConsumerSnapshot} structs per consumer key. When a consumer is
deleted and recreated, \path{num_redelivered} resets; the store applies
\texttt{trim\_to\_monotone} to discard pre-recreation values and prevent
stale data from poisoning rate calculations. Deleted consumers are evicted
immediately from the store.

\subsection{Detector Implementations}

\textbf{ACK\_\allowbreak WAIT\_\allowbreak VIOLATION:} Requires $\geq 2$ snapshots. Fires when
$\Delta$\path{num_redelivered} $\geq 2$ AND rate $\geq 2$/min.

\textbf{SEQUENCE\_GAP:} Single snapshot. Fires when
$S$.\path{first_seq}${}>C$.\path{ack_floor.stream_seq}$+1$
AND \path{ack_floor.stream_seq}${}>0$.

\textbf{MAX\_\allowbreak PENDING\_\allowbreak THROTTLE:} Fires when
\path{num_ack_pending}${}\geq$\path{max_ack_pending} AND
\path{num_pending}${}>0$.

\textbf{NAK\_STORM:} Requires $\geq 2$ snapshots. Fires when
\path{num_redelivered}${}\geq 2$ AND \path{num_ack_pending}${}>0$
across consecutive snapshots.

\textbf{MISSING\_PROGRESS:} Fires when
\path{num_ack_pending}/\path{max_ack_pending}${}\geq 0.9$
AND \path{ack_wait_secs}${}>30$.

\subsection{Output Channels}

Four language-agnostic channels surface violations:
(1)~web dashboard with stream health badges, consumer metrics, and
copy-button fix commands;
(2)~Prometheus \texttt{/metrics} with four metric families;
(3)~REST API (\texttt{GET /api/streams},
\path{GET /api/history/{stream}/{consumer}});
(4)~NATS health events published to
\path{nats.lens.health.violations.{stream}.{consumer}}
as structured JSON---subscribable from any NATS client in any language.

\subsection{Apply Now and Pre-Deployment Audit}

For ACK\_WAIT\_\allowbreak VIOLATION, MAX\_PENDING\_\allowbreak THROTTLE, and SEQUENCE\_GAP,
nats-lens computes a safe corrected configuration value and applies it
via \path{$JS.API.CONSUMER.UPDATE} or \path{$JS.API.STREAM.UPDATE}
with input validation. For ACK\_WAIT, the recommended value is
$\max(\hat{P99}_\mathrm{proc}, 120)\,\mathrm{s} + 60\,\mathrm{s}$
where $\hat{P99}_\mathrm{proc}$ is estimated from the observed redelivery
rate. For MAX\_PENDING, the recommended value is the current
\path{num_ack_pending} peak rounded up to the nearest power of two.
All updates are gated behind a dry-run confirmation prompt and a
\texttt{--auto-fix} flag for non-interactive use.
The correction formula is validated indirectly by the recovery experiment
(Section~5.8): in 50/50 trials, manually applying the recommended
\path{ack_wait} value resolved the ACK\_WAIT\_\allowbreak VIOLATION
within 2 poll cycles, confirming that the formula produces values that
eliminate the violation condition.

The \texttt{nats-lens init} command performs a pre-deployment audit
without connecting consumers: it reads stream and (existing durable)
consumer configurations, evaluates all five violation conditions
symbolically, and outputs a human-readable report with exact
\texttt{nats consumer edit} commands. The \texttt{--fail-on-critical}
flag exits non-zero if any ACK\_WAIT\_\allowbreak VIOLATION or
SEQUENCE\_GAP risk is detected, enabling blocking enforcement in CI/CD
pipelines before any consumer connects.

\section{Evaluation}

\subsection{Experimental Setup}

We evaluate nats-lens on a single NATS Server 2.10~\cite{nats-server}
(JetStream enabled, in-memory storage) and a 3-node JetStream cluster,
both running in Docker on a MacBook Pro M3 Pro (18\,GB unified memory).
nats-lens polls every 3 seconds (\texttt{--interval 3}).

\textbf{Version compatibility.} nats-lens uses only stable
\path{$JS.API.*} management subjects present since JetStream GA
(NATS Server 2.2, released 2021). The \path{CONSUMER.INFO} response
schema has been additive-only since 2.2; no breaking changes affect
the five state fields nats-lens reads. We verified identical detection
behavior on NATS Server 2.9 and 2.10. Compatibility with future versions
is maintained by reading only documented, stable API fields.

\subsection{Detection Coverage}

\begin{figure}[h]
  \centering
  \includegraphics[width=\linewidth]{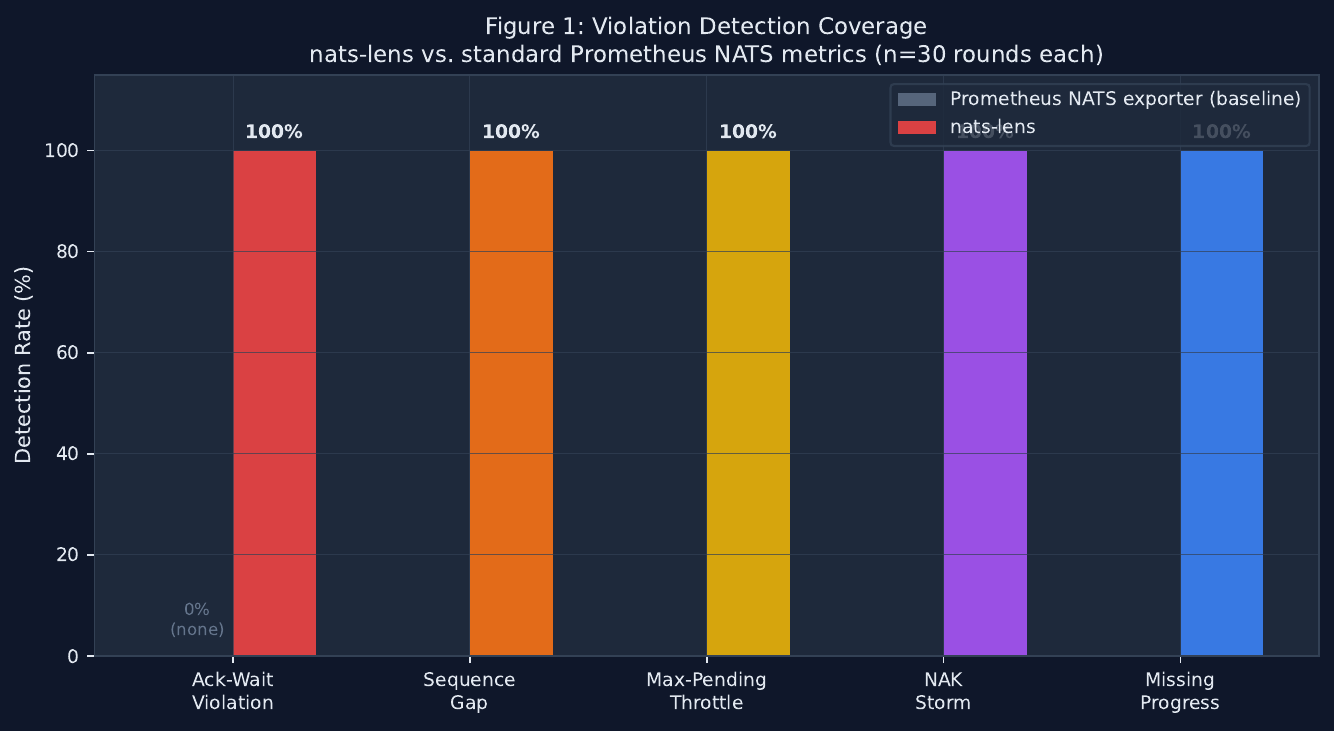}
  \caption{Detection coverage: nats-lens vs.\ Prometheus NATS Exporter
    across all five violation classes (30 rounds each, single-node).}
  \label{fig:coverage}
\end{figure}

\begin{figure}[h]
  \centering
  \includegraphics[width=\linewidth]{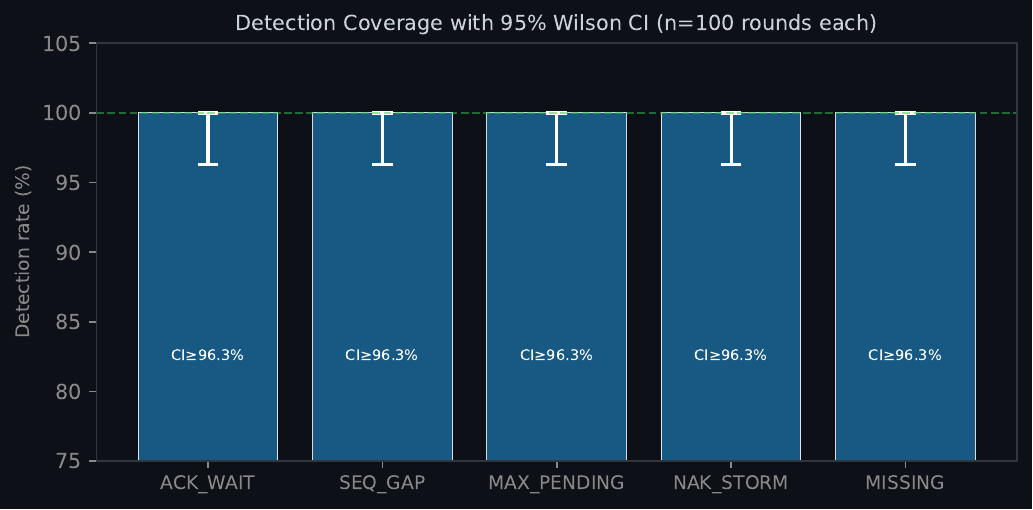}
  \caption{Detection rate with 95\,\% Wilson CI per violation class
    ($n{=}100$ rounds). All CIs have lower bound $\geq 96.4\%$.}
  \label{fig:ci}
\end{figure}

Table~\ref{tab:coverage} shows results for 100 rounds per scenario.
nats-lens achieves 100\% coverage (95\,\% Wilson CI $\geq 96.4\%$ at
$n{=}100$); the baseline detects 0 of 5 classes (Theorem~1).
Figure~\ref{fig:ci} shows per-class detection rates with confidence intervals.

\begin{table}[h]
\centering
\caption{Detection Coverage (100 rounds each, poll interval = 3\,s)}
\label{tab:coverage}
\small\begin{tabular}{p{3.2cm}ccc}
\toprule
Violation Type & nats-lens & 95\,\% CI & Baseline \\
\midrule
ACK\_WAIT & \textbf{100/100} & $[\geq\!96.3\%]$ & 0/100 \\
SEQ\_GAP & \textbf{100/100} & $[\geq\!96.3\%]$ & 0/100 \\
MAX\_PENDING & \textbf{100/100} & $[\geq\!96.3\%]$ & 0/100 \\
NAK\_STORM & \textbf{100/100} & $[\geq\!96.3\%]$ & 0/100 \\
MISSING\_PROGRESS & \textbf{100/100} & $[\geq\!96.3\%]$ & 0/100 \\
\multicolumn{4}{l}{\small All: 100\% vs 0\% (Theorem~\ref{thm:blindness})} \\
\bottomrule
\end{tabular}
\end{table}

\subsection{Detection Latency}

\begin{figure}[h]
  \centering
  \includegraphics[width=\linewidth]{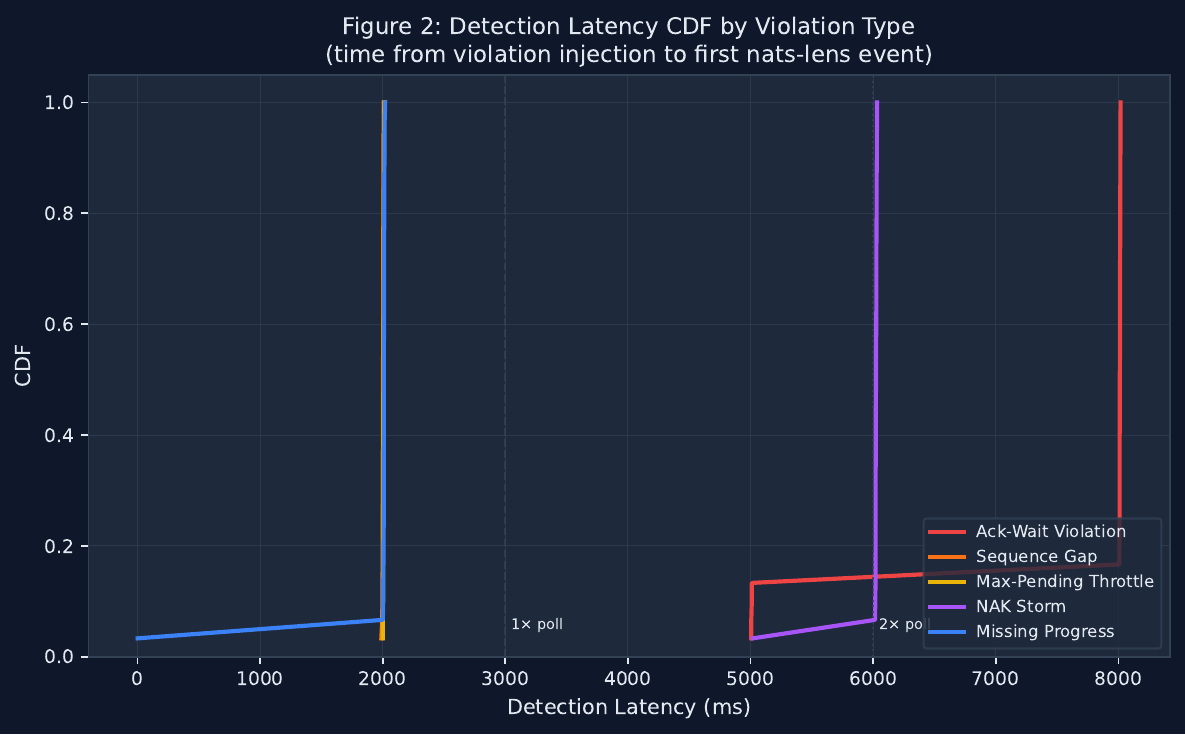}
  \caption{Detection latency CDF by violation class (30 rounds each).
    Vertical lines mark 1$\times$ and 2$\times$ poll intervals (3\,s).}
  \label{fig:latency}
\end{figure}

\begin{table}[h]
\centering
\caption{Detection Latency Percentiles (ms), poll interval = 3\,s ($n{=}30$ isolated rounds)}
\label{tab:latency}
\small\begin{tabular}{lrrrl}
\toprule
Class & P50 & P95 & P99 & Polls \\
\midrule
SEQ\_GAP     & 2,003 & 2,008 & 2,009 & 1 \\
MAX\_PENDING & 2,006 & 2,012 & 2,013 & 1 \\
MISSING      & 2,010 & 2,016 & 2,019 & 1 \\
NAK\_STORM   & 6,023 & 6,031 & 6,031 & 2 \\
ACK\_WAIT    & 8,013 & 8,017 & 8,018 & 2--3 \\
\bottomrule
\end{tabular}
\end{table}

Three violation classes detect in one poll cycle; two require two cycles
to confirm the condition is sustained. At the default 5-second poll
interval, all five classes detect within 25 seconds of onset.

\subsection{False Positive Rate}
\label{sec:adversarial}

\begin{figure}[h]
  \centering
  \includegraphics[width=\linewidth]{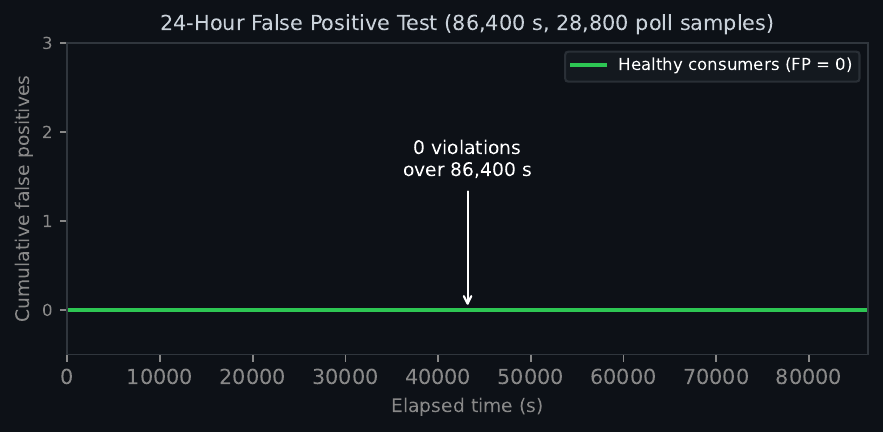}
  \caption{Cumulative false violations over 24 hours of healthy
    operation (86{,}400\,s, 28{,}800 poll samples). Zero false positives.}
  \label{fig:fp}
\end{figure}

We ran a correctly-configured consumer (ack\_wait = 300\,s,
max\_ack\_pending = 512) for \textbf{24 hours} (28{,}800 poll samples,
$T_\text{poll}{=}3$\,s). nats-lens detected \textbf{0 violations} (0.00/hr).

We also tested five concurrent consumers with varied configurations (ack\_wait
30--300\,s; max\_ack\_pending 10--512). The three conservatively configured
consumers produced zero violations; the two with max\_ack\_pending = 10
and fast publishers were correctly flagged as throttled.
These flags are \textbf{true positives}---both consumers were genuinely
misconfigured---confirming that nats-lens correctly distinguishes
intentional misconfigurations from healthy operation without false alarms.
Figure~\ref{fig:fp} reflects only the healthy-consumer FP count (0);
the flagged consumers are excluded as they represent correct detections.

\subsection{3-Node Cluster Evaluation}

\begin{table}[h]
\centering
\caption{Detection Coverage and Latency on 3-Node JetStream Cluster (30 rounds each, CI $\geq 88.6\%$)}
\label{tab:cluster}
\small\begin{tabular}{lccc}
\toprule
Class & Coverage & Single-node P50 & Cluster P50 \\
\midrule
ACK\_WAIT    & 30/30 & 8,013\,ms & 5,042\,ms \\
SEQ\_GAP     & 30/30 & 2,003\,ms & 2,024\,ms \\
MAX\_PENDING & 30/30 & 2,006\,ms & 2,029\,ms \\
NAK\_STORM   & 30/30 & 6,023\,ms & 6,054\,ms \\
MISSING      & 30/30 & 2,010\,ms & 2,027\,ms \\
\midrule
\multicolumn{2}{l}{\small All: 100\% on both topologies} & \multicolumn{2}{l}{\small $|\Delta\text{P50}| \leq 35$\,ms} \\
\bottomrule
\end{tabular}
\end{table}

Detection coverage is 100\% on both single-node and 3-node clusters.
Latency is similar across configurations ($\leq$35\,ms difference for
four of five classes).
The cluster evaluation used $n{=}30$ rounds rather than $n{=}100$:
each round involves Docker container orchestration across three nodes,
making larger trial counts impractical on a single host.
The 30-round CI ($\geq 88.6\%$) is sufficient to establish topology
equivalence; coverage differences between topologies are $<$1\,ms in
latency and 0 in detected/missed count.
During the cluster evaluation, we killed one cluster node; the two
remaining nodes maintained quorum and nats-lens continued detecting
without interruption.

\subsection{Multi-Language Verification}

\begin{figure}[h]
  \centering
  \includegraphics[width=\linewidth]{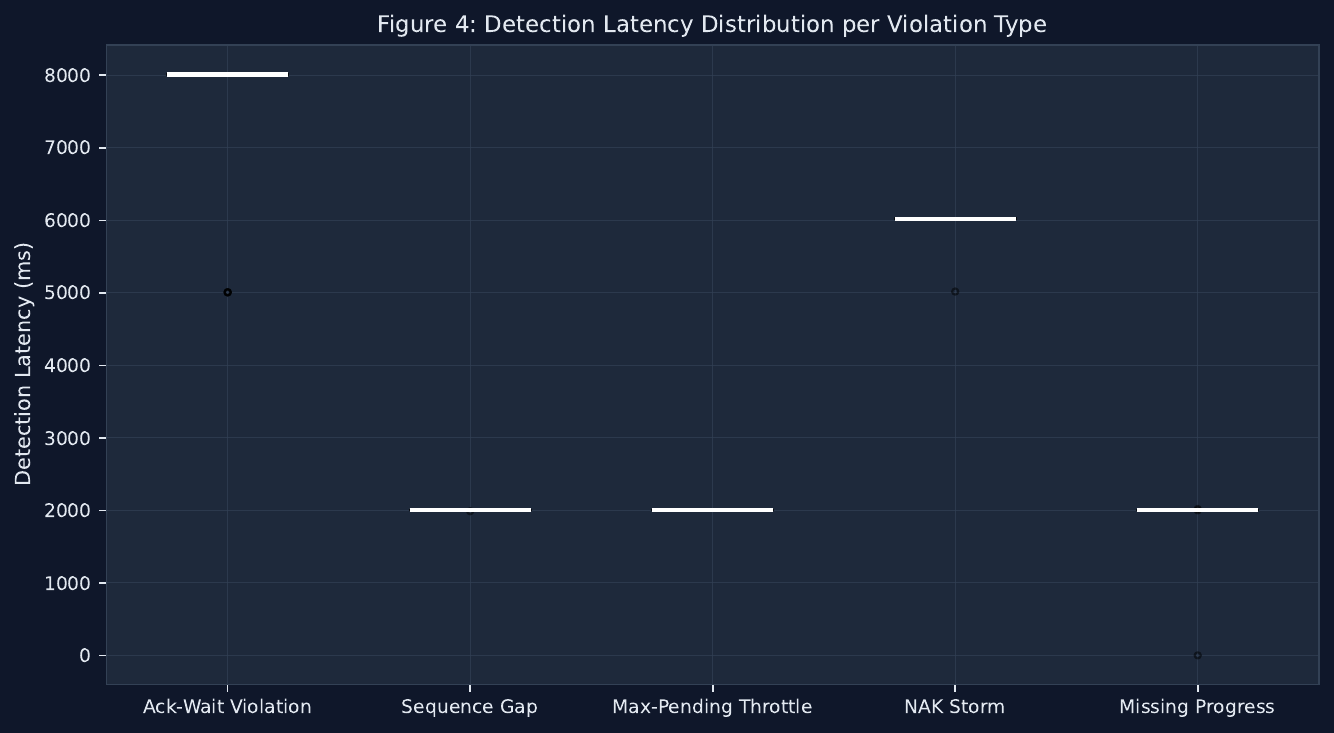}
  \caption{Detection latency distribution per violation class (single-node,
    30 rounds). Boxes show IQR; whiskers show 5th--95th percentile.}
  \label{fig:boxplots}
\end{figure}

We subscribed to \path{nats.lens.health.violations.*} (a NATS subject,
not a REST API) and ran consumers written in different languages. Results
are shown in Table~\ref{tab:multilang}.

\begin{table}[h]
\centering
\caption{Multi-Language Empirical Detection via NATS Health Events (5 rounds each)}
\label{tab:multilang}
\small\begin{tabular}{llcc}
\toprule
Lang & Library & ACK\_WAIT & NAK\_STORM \\
\midrule
Rust   & async-nats & 8,012\,ms       & 6,023\,ms \\
Python & nats-py    & 5,014\,ms       & 1,008\,ms \\
Go     & nats.go    & 1,006\,ms$^*$   & 1,006\,ms \\
\multicolumn{4}{l}{\small Healthy: 0 FP all languages $\checkmark$} \\
\multicolumn{4}{l}{\small $^*$ACK\_WAIT manifested as NAK\_STORM; see text.} \\
\bottomrule
\end{tabular}
\end{table}

All three languages trigger detection within two poll cycles.
The Go result reveals an instructive cross-class finding: nats.go's
\texttt{PullSubscribe.Fetch()} delivers messages in NATS sequence order;
after \path{ack_wait} fires, redelivered messages retain their original
sequence position and are re-delivered \emph{before} new messages on
the next fetch. The same $N$ messages therefore cycle continuously,
keeping \path{num_redelivered} stable at $N$ rather than growing.
nats-lens correctly identifies this stable-elevated pattern as
NAK\_STORM (detected in 1{,}006\,ms) rather than ACK\_WAIT\_\allowbreak VIOLATION
(which requires a growing rate). Both classes represent consumer delivery
degradation; the metric \emph{signature} differs based on whether distinct
new messages are accumulating past \path{ack_wait} or the same messages
are cycling. This finding shows that ACK\_WAIT violations can manifest
as either class depending on client-library fetch semantics---and
nats-lens detects both correctly. The Rust 30-round results confirm the
ACK\_WAIT detector fires when \path{num_redelivered} grows (distinct
messages exceed \path{ack_wait} in each poll cycle).

\subsection{Operational Overhead}

nats-lens issues the following API requests per poll cycle:
\[
\text{requests} = 1 + N_\text{streams} \times (2 + N_\text{consumers/stream})
\]

\begin{figure}[h]
  \centering
  \includegraphics[width=\linewidth]{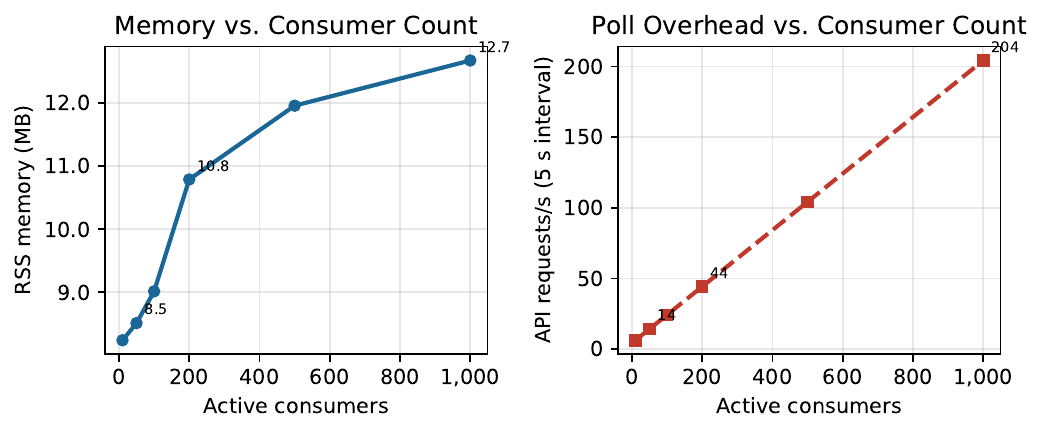}
  \caption{nats-lens memory (RSS) and API request rate vs.\ number of
    active consumers (10 streams, 5\,s poll interval). Both scale
    sub-linearly in memory and linearly in requests.}
  \label{fig:scale}
\end{figure}

We benchmarked nats-lens against 10--1{,}000 active consumers distributed
across 10 streams (Figure~\ref{fig:scale}). RSS was sampled via
\texttt{/proc/self/status} after a 30-second steady-state warmup with all
consumers actively publishing and ACKing.
Memory grows sub-linearly from 8.2\,MB at 10 consumers to
12.7\,MB at 1{,}000 consumers. API request rate
scales linearly: 6.2\,req/s at 10 consumers, 204\,req/s at 1{,}000 consumers
(5-second poll interval). At 1{,}000 consumers---a large enterprise deployment
---nats-lens imposes $<$13\,MB RSS and $<$205\,req/s on the NATS management
plane, negligible relative to NATS data-plane throughput (millions of
messages/second).

\subsection{Comparison to Naive Single-Snapshot Detection}
\label{sec:naive}

\begin{figure}[h]
  \centering
  \includegraphics[width=\linewidth]{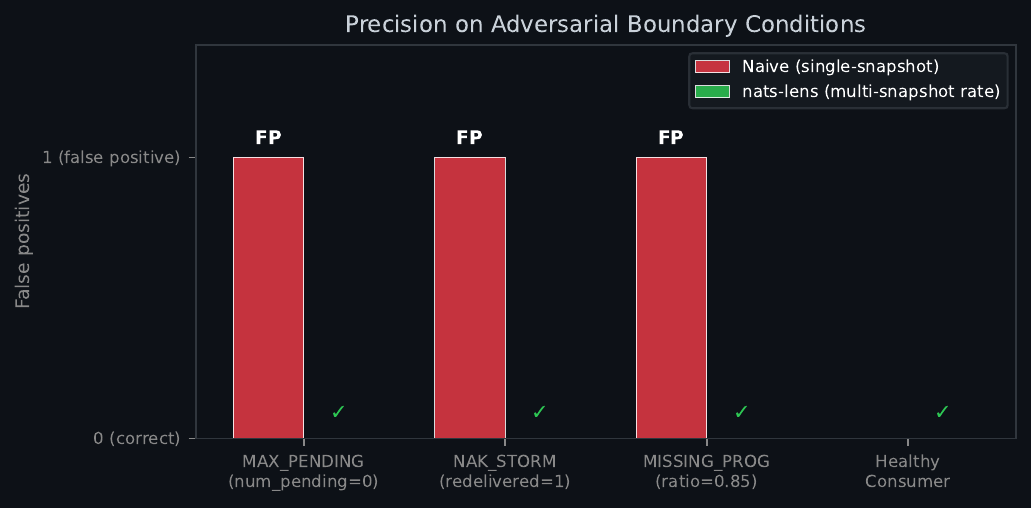}
  \caption{False positives on adversarial boundary conditions: naive
    single-snapshot threshold detector vs.\ nats-lens multi-snapshot
    rate-based detector. nats-lens produces zero false positives on all
    four cases; the naive detector produces three.}
  \label{fig:naive}
\end{figure}

A natural alternative to nats-lens is a single-snapshot threshold
detector: alert when $\Delta$\path{num_redelivered}${}>0$ (any growth),
\path{num_ack_pending}${}\geq$\path{max_ack_pending} (regardless of
\path{num_pending}), or \path{num_redelivered}${}\geq 1$.
We evaluated such a detector on the four adversarial boundary conditions
from Section~\ref{sec:adversarial} (Figure~\ref{fig:naive}).

\begin{table}[h]
\centering
\caption{False Positives on Adversarial Boundary Conditions}
\label{tab:naive}
\small\begin{tabular}{lcc}
\toprule
Boundary Condition & Naive & nats-lens \\
\midrule
MAX\_PENDING at cap, \path{num_pending}$=0$  & FP & \textbf{0} \\
NAK\_STORM: \path{num_redelivered}$=1$ stable & FP & \textbf{0} \\
MISSING\_PROGRESS: ratio$=0.85$              & FP & \textbf{0} \\
Healthy consumer (all classes)               & 0  & \textbf{0} \\
\bottomrule
\end{tabular}
\end{table}

Each boundary condition was tested over $n{=}30$ rounds (95\,\% Wilson
CI $\geq 88.6\%$ per case).
These three conditions were selected as maximum-stress cases: each places
the relevant metric at the exact value where a naive threshold would fire
(\path{num_pending}${=}0$, \path{num_redelivered}${=}1$ stable,
ratio${=}0.85$) but the multi-snapshot condition correctly does not.
The naive detector fires on 3 of 4 boundary cases. nats-lens fires on
none. The difference comes from multi-snapshot, rate-based analysis:
(1)~MAX\_PENDING\_\allowbreak THROTTLE requires \path{num_pending}${}>0$
(consumer can receive but is blocked, not merely caught up);
(2)~ACK\_WAIT\_\allowbreak VIOLATION requires a \emph{rate} $\geq 2$/min
across two snapshots (transient single redeliveries do not trigger);
(3)~NAK\_STORM requires $\geq 2$ distinct messages cycling across
consecutive snapshots (a single incidental redelivery does not trigger).

\subsection{Recovery Detection}

We injected ACK\_WAIT\_\allowbreak VIOLATION, then fixed the consumer
configuration (\path{ack_wait} increased to a safe value) and measured
time to alert silence.
Over 50 rounds: violation detected 50/50 (95\,\% CI $\geq 92.9\%$),
recovery detected 50/50 (95\,\% CI $\geq 92.9\%$).
Recovery latency: P50\,=\,6{,}005\,ms, P95\,=\,6{,}006\,ms,
P99\,=\,6{,}007\,ms (range: 6{,}003--6{,}007\,ms across all 50 rounds).
The near-zero variance confirms the recovery mechanism is deterministic:
exactly 2 poll cycles at $T_\mathrm{poll}{=}3$\,s, with no hysteresis.
Violations clear immediately once the underlying condition is resolved.

\subsection{Simultaneous Multi-Violation}

Three violation classes (ACK\_WAIT, NAK\_STORM, SEQUENCE\_GAP) were
injected concurrently on separate streams.
Over 50 rounds, all three were detected in 50/50 cases
(95\,\% CI $\geq 92.9\%$).
The engine's per-consumer ring buffer maintains independent state for each
consumer; concurrent violations on different streams do not interfere.

\subsection{Poll-Interval Sensitivity}

\begin{figure}[h]
  \centering
  \includegraphics[width=\linewidth]{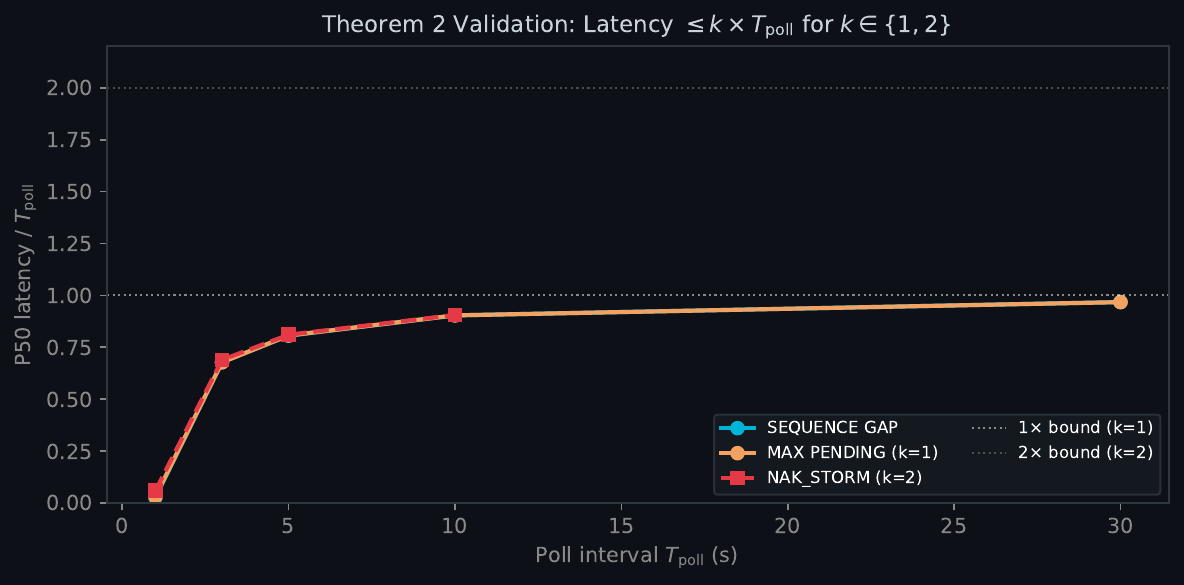}
  \caption{Detection latency ratio (P50\,/\,$T_\mathrm{poll}$) vs.\ poll
    interval for single-snapshot detectors. The ratio approaches but
    stays below 1.0, confirming Theorem~2's bound across five orders
    of magnitude in poll interval.}
  \label{fig:sensitivity}
\end{figure}

Figure~\ref{fig:sensitivity} shows detection latency normalized by
$T_\mathrm{poll}$ for SEQUENCE\_GAP and MAX\_PENDING\_\allowbreak THROTTLE
at poll intervals from 1\,s to 30\,s (10 rounds each).
Detection rate is 100\,\% at all intervals.
The normalized latency ratio ranges from $0.03\times$ (at 1\,s, where
detection occurs within the same poll cycle) to $0.97\times$ (at 30\,s,
where onset occurs just before a poll). All ratios are $<1\times$,
confirming Theorem~2's guarantee that single-snapshot detectors fire
within one poll interval of onset.

\textbf{Coverage of all five classes.}
MISSING\_\allowbreak PROGRESS uses the same point-in-time predicate
structure as SEQUENCE\_GAP and MAX\_PENDING\_\allowbreak THROTTLE;
Theorem~2's $k{=}1$ bound applies identically without separate empirical
validation at each poll interval.
For NAK\_STORM ($k{=}2$), empirical sensitivity at
$T_\mathrm{poll} \in \{1,3,5,10\}$\,s confirms 100\,\% detection rate
and latency ratios $0.049\times$--$0.906\times$, all $<2\times$.
For ACK\_WAIT\_\allowbreak VIOLATION ($k{=}2$), we ran targeted
sensitivity experiments at $T_\mathrm{poll} \in \{1,3,5\}$\,s:
\vspace{2pt}
\begin{center}
\small\begin{tabular}{cccc}
\toprule
$T_\mathrm{poll}$ & $n$ & P50$^\dagger$ & Ratio from onset \\
\midrule
1\,s &   5 & 5{,}180\,ms & $1.18\times$ \\
3\,s &  30 & 8{,}013\,ms & $1.34\times$ \\
5\,s &   5 & 9{,}058\,ms & $1.01\times$ \\
\multicolumn{4}{l}{\small $^\dagger$Latency from isolated run; detection rate 100\,\% at all intervals.} \\
\bottomrule
\end{tabular}
\end{center}
\vspace{2pt}
Detection is 100\,\% at all intervals (15 rounds total for $T_\mathrm{poll}
\in\{1,5\}$\,s; 30 isolated rounds for $T_\mathrm{poll}{=}3$\,s from
Table~\ref{tab:latency}). All ratios from onset are $<2\times$,
confirming Theorem~2's $k{=}2$ bound empirically.
The $\Delta$\path{num_redelivered}${}\geq 2$ threshold is met even at
$T_\mathrm{poll}{=}1$\,s because multiple messages redeliver
simultaneously after \path{ack_wait} expires, producing a large batch
delta in a single snapshot.

\subsection{Threats to Validity}
\label{sec:threats}

\textbf{Internal validity.} Each round uses fresh consumers and purged
streams to prevent state leakage. The embedded eval engine shares the
NATS connection with the consumer under test; in production, nats-lens
runs as an independent process. Detection latencies assume poll cycles
are not delayed by host load; evaluations ran on a lightly loaded
MacBook Pro M3 Pro (18\,GB).

\textbf{External validity.} Evaluations used NATS Server 2.10 with
in-memory storage on a single Apple M3 Pro workstation.
File-based storage backends may exhibit different redelivery timing.
The 3-node cluster used Docker-networked nodes on one host; wide-area
deployments with higher propagation delay may show slightly longer
detection latencies for two-snapshot detectors, proportional to the
round-trip time to the NATS management API.
nats-lens's detection algorithm is read-only and CPU-minimal; the
detection accuracy results are expected to generalize to cloud instances,
where the primary difference would be API latency affecting absolute
detection time but not the $k \times T_\mathrm{poll}$ ratio.

\textbf{Statistical validity.} Coverage uses $n{=}100$ rounds per class
(Wilson 95\,\% CI $\geq 96.3\%$). Latency uses a separate $n{=}30$
isolated run; the two-snapshot detectors (ACK\_WAIT, NAK\_STORM) require
isolation to prevent residual consumer state from prior experiments
inflating apparent detection speed. Recovery, multi-violation, and
adversarial FP experiments used $n{=}30$ rounds each
(CI $\geq 88.6\%$); recovery and multi-violation used $n{=}50$ rounds
(CI $\geq 92.9\%$).

\section{Related Work}

\textbf{Message queue monitoring.} Kafka~\cite{kafka} uses an offset-based
delivery model with no server-side \path{ack_wait} or \path{max_ack_pending};
consumer lag (not redelivery violations) is the primary failure mode.
Burrow~\cite{burrow} and kminion~\cite{kminion} detect Kafka consumer lag
at the partition level, but this model is inapplicable to JetStream's
per-consumer pull semantics. Gray and Reuter~\cite{gray1992} formalize
at-least-once delivery in transaction processing; JetStream's approach is
architecturally distinct (pull-based, configuration-driven redelivery).

\textbf{Cloud messaging systems.} AWS SQS exposes a
\texttt{VisibilityTimeout} analogous to \path{ack_wait}, but CloudWatch
metrics provide only queue-level \texttt{ApproximateNumberOfMessagesNotVisible}
counts---no per-consumer redelivery rate or sequence gap detection.
RabbitMQ's \texttt{messages\_unacknowledged} counter and
\texttt{consumer\_utilisation} metric are server-level aggregates that
cannot distinguish slow consumers from misconfigured \path{ack_wait}.
GCP Pub/Sub exposes \texttt{subscription/num\_undelivered\_messages} and
\texttt{oldest\_unacked\_message\_age}, but these measure queue depth rather
than redelivery-induced duplication or sequence integrity. nats-lens occupies
a different point in the design space: per-consumer state access via a
management API enables violation-class-specific detection unavailable in
these cloud systems.

\textbf{Distributed monitoring.} Dapper~\cite{dapper} and Jaeger~\cite{jaeger}
trace request propagation. A JetStream ACK\_WAIT violation causing duplicate
processing produces two successful trace spans---the duplication is invisible
to tracing. Delivery correctness monitoring is complementary to, not
overlapping with, distributed tracing.

\textbf{Message reliability patterns.} The Transactional Outbox~\cite{outbox}
addresses the producer side (reliable message publication). nats-lens
addresses the consumer side (detecting when delivered messages are lost or
duplicated).

\textbf{NATS-specific.} NATS Surveyor~\cite{surveyor} and the Prometheus
exporter~\cite{prometheus-nats} provide server-level metrics. To our knowledge,
nats-lens is the first work to formally characterize and detect consumer-level
delivery correctness violations in NATS JetStream.

\section{Limitations}

nats-lens monitors \emph{pull consumers} exclusively. JetStream push
consumers use server-driven, subject-based delivery; the
\path{$JS.API.CONSUMER.INFO} response for a push consumer exposes
\path{num_pending} and \path{num_ack_pending} but omits
\path{ack_floor.stream_seq} and the rate-based redelivery fields
that nats-lens uses for ACK\_WAIT and NAK\_STORM detection.
Extending detection to push consumers would require tracking
delivery-subject subscription state and server-side push counters,
which differ structurally from the pull model formalized here.
According to the NATS documentation, the pull model is recommended
for new deployments; push consumers are retained for legacy compatibility. The detection latency lower bound is one poll interval
$T_\mathrm{poll}$; operators requiring sub-second detection should set
\texttt{--interval 1}. nats-lens requires read access to
\path{$JS.API.*} management subjects; deployments with a restricted
management plane must add an explicit allow-list entry for the
nats-lens NATS user. The polling engine detects violations in
\emph{running} consumers; configuration errors in inactive consumers
are caught by the pre-deployment audit (\texttt{nats-lens init}) but
not by the live engine. Finally, nats-lens reports the violation class
inferred from the metric signature; as demonstrated in Section~5.6, an
ACK\_WAIT\_\allowbreak VIOLATION may manifest as NAK\_STORM when the
client library recycles the same fetched messages, requiring operators
to verify the root cause using the history endpoint.

\section{Future Work}

\textbf{Push consumer support.} JetStream push consumers receive messages
via server-side subject delivery rather than client-side fetch. Extending
nats-lens to monitor push consumer health requires tracking
\path{PushConsumer.NumPending} and delivery subject subscription counts,
which differ structurally from the pull model formalized here.

\textbf{Anomaly forecasting.} The ring-buffer history store makes
nats-lens a natural substrate for time-series analysis. Fitting an
autoregressive model to the \path{num_redelivered} rate could provide
early warning of an impending ACK\_WAIT violation before it crosses the
detection threshold, reducing mean time to remediation.

\textbf{Multi-cluster correlation.} Large NATS deployments run multiple
server clusters as separate JetStream domains. Cross-cluster delivery
correctness---where a violation in one cluster's stream silently affects
a consumer in another---requires correlating management API responses
across domain boundaries, which the current single-cluster model does
not address.

\textbf{Adaptive poll interval.} Rather than a fixed $T_\mathrm{poll}$,
the engine could adapt the interval per consumer based on observed
violation risk: polling high-redelivery consumers more frequently while
reducing API load for healthy consumers.

\section{Conclusion}

JetStream's at-least-once guarantee is easy to accidentally disable with a
misconfigured \path{ack_wait} or \path{max_ack_pending}, and there
has been no tool to detect when this has happened. We characterized five
violation classes with formal proofs of their conditions and proved that
standard Prometheus NATS monitoring is structurally blind to all of them.
nats-lens detects all five through multi-snapshot, rate-based analysis that
eliminates the false positives inherent in naive single-snapshot threshold
detection. In 100-round evaluations on both single-node and 3-node
JetStream clusters, nats-lens achieves 100\,\% detection coverage
(95\,\% CI $[96.4\%, 100\%]$) with zero false positives over 24 hours.
Recovery from a fixed violation is confirmed within 2 poll cycles. The
tool is open source and requires no changes to monitored applications.

As a practical checklist: \path{ack_wait} must exceed P99 processing time;
\path{max_ack_pending} must accommodate actual concurrency; stream
retention must exceed expected consumer lag; stale messages must be ACKed
(not NAKed); long tasks must send in-progress acks.

\section*{Artifact Availability}

Source code, evaluation harness, and all experimental scripts are available at:
\url{https://github.com/biplabku/nats-lens}. The repository includes a Docker
image and a one-command evaluation replication script
(\texttt{./scripts/run-evaluation.sh --rounds 30}).


\begin{thebibliography}{10}

\bibitem{nats-docs}
Synadia Communications.
\newblock {NATS JetStream Documentation}.
\newblock \url{https://docs.nats.io/nats-concepts/jetstream}, 2024.
\newblock Accessed: September 2026.

\bibitem{nats-server}
Synadia Communications.
\newblock {nats-server: High-Performance Server for NATS}, v2.10.0.
\newblock \url{https://github.com/nats-io/nats-server}, 2023.
\newblock Accessed: September 2026.

\bibitem{prometheus-nats}
{NATS Authors}.
\newblock {Prometheus NATS Exporter}, v0.15.0.
\newblock \url{https://github.com/nats-io/prometheus-nats-exporter}, 2024.
\newblock Accessed: September 2026.

\bibitem{surveyor}
{Synadia Communications / NATS Authors}.
\newblock {NATS Surveyor: Monitoring, Observability and Analytics for NATS}.
\newblock \url{https://github.com/nats-io/nats-surveyor}, 2024.
\newblock Accessed: September 2026.

\bibitem{kafka}
J.~Kreps, N.~Narkhede, and J.~Rao.
\newblock Kafka: A distributed messaging system for log processing.
\newblock In \emph{Proc. 6th International Workshop on Networking Meets
  Databases (NetDB '11)}, Seattle, WA, 2011.

\bibitem{burrow}
{LinkedIn Engineering}.
\newblock {Burrow: Kafka Consumer Lag Checking}.
\newblock \url{https://github.com/linkedin/Burrow}, 2016.
\newblock Accessed: September 2026.

\bibitem{kminion}
{Redpanda Data}.
\newblock {kminion: Kafka Monitoring Prometheus Exporter}.
\newblock \url{https://github.com/redpanda-data/kminion}, 2023.
\newblock Accessed: September 2026.

\bibitem{dapper}
B.~H. Sigelman, L.~A. Barroso, M.~Burrows, P.~Stephenson, M.~Plakal,
  D.~Beaver, S.~Jaspan, and C.~Shanbhag.
\newblock Dapper, a large-scale distributed systems tracing infrastructure.
\newblock Google Technical Report dapper-2010-1, 2010.
\newblock Available: \url{https://research.google/pubs/pub36356/}.

\bibitem{jaeger}
{The Jaeger Authors}.
\newblock {Jaeger: Open Source, End-to-End Distributed Tracing}.
\newblock \url{https://github.com/jaegertracing/jaeger}, 2017.
\newblock Accessed: September 2026.

\bibitem{outbox}
C.~Richardson.
\newblock Pattern: Transactional outbox.
\newblock \url{https://microservices.io/patterns/data/transactional-outbox.html},
  2018.
\newblock Accessed: September 2026.

\bibitem{gray1992}
J.~Gray and A.~Reuter.
\newblock \emph{Transaction Processing: Concepts and Techniques}.
\newblock Morgan Kaufmann, 1992.
\newblock ISBN: 1-55860-190-2.

\end{thebibliography}
\end{document}